# Comparison of JET main chamber erosion with dust collected in the divertor

A. Widdowson[a*], C.F. Ayres[a], S. Booth[a], J.P. Coad[b], A. Hakola[b], K. Heinola[c], D. Ivanova[d], S. Koivuranta[b], J. Likonen[b], M. Mayer[e], M. Stamp[a] and JET-EFDA contributors[f]

*JET-EFDA, Culham Science Centre, OX14 3DB, Abingdon, UK*

[a]*EURATOM/CCFE Fusion Association, Culham Science Centre, Abingdon, OX14 3DB, UK*

[b]*Association EURATOM-TEKES, VTT, PO Box 1000, 02044 VTT, Espoo, Finland*

[c]*Association EURATOM-TEKES, University of Helsinki, PO Box 64, 00560 Helsinki, Finland*

[d]*Laboratory, Royal Institute of Technology, Association EURATOM-VR, 100 44 Stockholm, Sweden*

[e]*Max-Planck-Institut für Plasmaphysik, EURATOM Association, 85748 Garching, Germany*

[f]*See the Appendix of F Romanelli et al., Proc 23rd IAEA Fusion Energy Conference 2010 Daejon, Korea*

**Abstract**

A complete global balance for carbon in JET requires knowledge of the net erosion in the main chamber, net deposition in the divertor and the amount of dust and flakes collecting in the divertor region. This paper describes a number of measurements on aspects of this global picture. Profiler measurements and cross section microscopy on tiles that were removed in the 2009 JET intervention are used to evaluate the net erosion in the main chamber and net deposition in the divertor. In addition the mass of dust and flakes collected from the JET divertor during the same intervention is also reported and included as part of the balance. Spectroscopic measurements of carbon erosion from the main chamber are presented and compared with the erosion measurements for the main chamber.

*Corresponding author address:* EURATOM/CCFE Fusion Association, Culham Science Centre, Abingdon, OX14 3DB, UK

*Corresponding author E-mail:* anna.widdowson@ccfe.ac.uk

*Presenting author:* Dr A. Widdowson

*Presenting author email:* anna.widdowson@ccfe.ac.uk

1. Introduction

2010 marked the end of an era for the JET vessel and subsequent *post mortem* analysis of tiles. Since the installation of the divertor in 1994, JET has operated as an *all carbon* machine in that all surfaces with direct interaction with the confined plasma were made from carbon, either as graphite or latterly carbon fibre composite (CFC). Just over half of the vessel was covered in carbon tiles and the remaining uncovered area was the inconel vacuum vessel. In 2010 JET was converted to an *all metal* device such that all surfaces interacting with the plasma are now beryllium and tungsten. This new configuration is known as the ITER-like Wall (ILW) since it is designed to demonstrate the differences in transport and hydrogen isotope retention between the two scenarios and to help predict the behaviours of ITER in these respects. In order to complete the transition from an *all carbon* to *all metal* wall all CFC tiles were removed and replaced with Be, Be-coated inconel or W coated CFC tiles in the main chamber and W coated CFC with one row of solid W tiles in the divertor. As with other JET interventions a set of tiles removed from the vessel have been made available for analysis. The complete refurbishment also provided a unique opportunity to collect dust and flakes found in the divertor as all the divertor carriers were removed from the vessel.

There are many references on JET and other machines discussing material migration based on *post mortem* analysis of tiles and passive diagnostics, however many do not bring data together to give an overall global picture. An overview on erosion/deposition studies on JT60U [1] concludes that net deposition exceeds net erosion in the divertor which in turn implies that the carbon source for deposits comes predominantly from the main chamber, although evaluation of the main chamber source is not discussed. Data for TEXTOR [2] shows a balance with limiter plasmas with the main source of erosion being the toroidal belt limiter and the main deposition areas being the toroidal belt limiter and other "obstacles" in the scrape off layer. Other particle balance exercises have been presented, for example the Deuterium Inventory in Tore Supra (DITS) programme which extensively reported erosion/deposition from *post mortem* analysis of tiles in Tore Supra to investigate fuel retention [3].

In this paper the mass of carbon eroded in the main chamber of JET is compared with the mass of carbon found in the form of deposits and dust/flakes in the divertor. In addition an estimate of the carbon source is determined from spectroscopy of the CIII line in the main chamber throughout the last operating period (2007 - 2009). The results give an insight into the scale of carbon migrating around the vessel during a JET operating period typically lasting > 100 000 seconds. A similar global carbon balance for JET was presented for the 1999 - 2001 Mark II Bas Box Divertor configuration [4] where errors in the balance between the main chamber source determined from CIII spectroscopy and the deposits found on divertor was within a factor of two.

2. **Experimental Details**

The erosion and deposition of inner wall guard limiter (IWGL) tiles, outer poloidal limiter (OPL) tiles and dump plate tiles from the main chamber and tiles constituting a poloidal divertor cross section has been measured by profiling the surfaces of a set of tiles before and after exposure in the vessel. Figure 1 shows the poloidal location of the tiles analysed for this paper. Each IWGL location is made up from a pair of tiles; a left hand and a right hand tile (looking to the centre of the machine). Each OPL tile is a tile pair placed one on top of the other; a top tile and a bottom tile. The profiler consists of a X-Y table and a Z probe. The tile to be profiled is mounted onto the X-Y table which is moved using two stepper motors to a series of specified (X,Y) co-ordinates forming a grid. At each grid point the Z probe is extended to touch the surface of the tile thus recording a relative value for the height.

The comparison of the grid measurements on a tile before and after installation in the JET vessel gives the change in the surface profile, i.e. the erosion and deposition on the tile surface. The profiler itself provides repeatable measurements to within a few microns, however errors can arise if the tile is not repositioned on the X-Y table accurately. On horizontal surfaces this error is minimal, however where the tile surface slopes the errors increase. From simple trigonometry the errors in the height is 0.18 μm per micron misalignment on a surface at 10º to the horizontal. Based on this misalignment error tile profiling provides an assessment of erosion on a scale >10 μm. From the profiler results the volume of deposited and eroded material is determined. To convert volume to mass a density of 1 g/cm$^3$ has been used for the co-deposit [5] [6] and a density of 1.65 g/cm$^3$ has been used for erosion from the CFC tiles - this is an average value for batches of Dunlop CFC material. The mass of eroded/deposited carbon is scaled with the number of similar tiles found in the vessel to give values for the whole vessel.

In addition to the results from the tile profiling, information on deposition has also been obtained from the optical microscopy of cross sections of cores cut from tiles. This has enabled the deposition results from the tiles profiled to be bench-marked against the core samples thus providing a calibration and giving confidence in the erosion assessment provided from the profiler results. The comparison of cross section optical microscopy with profiler results is ongoing and has so far only been completed for a sub-set of profiler results presented in this paper.

An alternative method for the evaluation of erosion is to use a marker coating on the tile and to access the thinning of the coating after exposure in the vessel. Typically a marker coating is 10 μm thick to enable the initial thickness of the coating to be analysed by ion beam techniques. However the thickness of the coating limits the amount of erosion that can be measured. If the marker coating is completely removed then the coating thickness cannot be analysed and thus the total erosion cannot be evaluated. For these reasons ion beam analysis of tiles with marker coatings are useful where erosion is expected to be less than 10 μm. For example this technique has been used to determine erosion from inner wall cladding (IWC) tiles [7] exposed in JET from 2005-2009.

During the 2012 shutdown dust/flakes were collected from the divertor region of JET using a vacuum cleaner and cyclone adapted for use by remote handling. A cyclone pot was installed at the bottom of the cyclone to capture the dust/flake sample during vacuuming. Six different vacuum samples from different poloidal regions of interest in the divertor were collected into six cyclone pots. The regions are indicated in Figure 2. In order of collection the regions are; the outer vertical divertor tiles and above (tiles B, C, 7, 8 and Load Bearing Tile), the inner divertor tiles (High Field Gap Closure tile, tile 1, 3), the inner and outer divertor carrier ribs, the outer floor tiles (Tile 6), inner floor tiles (Tile 4), the inner and outer louvre regions. The cyclone pots were weighed before and after collection to determine the mass of dust/flakes collected. It was necessary to compensate for masses obtained in the cyclone pots in several ways, for example: (i) some dust may be lodged in the hose or cyclone or may have bypassed the cyclone (due to the mass of the particle) and be captured in the dust bag; (ii) 11/12ths of the divertor area was vacuumed in 5 out of 6 samples, only the inner and outer louvre region was vacuumed in its entirety; (iii) some surfaces have been vacuumed in previous interventions, i.e. there is a range of histories for different surfaces.

In order to compensate for (i) the tritium off-gas rates for each of the six dust samples collected into the cyclone pots was measured. From this the *specific* off-gas rate (Bq / day / g) for each of the six dust/flake samples was determined. Using the specific off-gas rate for the dust/flake samples it was possible to determine the mass of material trapped in the hose, cyclone and dust bag associated with the sample from off-gassing assessments of this equipment, thus providing an additional contribution to the overall mass collected. The specific off-gas rates varied from < 1 GBq/g for the sample from the vertical inner divertor tiles (HFGC, Tile 1 and Tile 3) to 62 G Bq/g for the base tiles at the inner divertor (Tile 4). Based on these values trapped dust/flakes in the vacuum hose and cyclone was typically <1% of the total mass collected in that sample. In comparison the mass of dust/flakes in the vacuum bag was considerably higher, ranging from <10% to >30% of the dust collected in the corresponding sample pot.

During the vacuuming of the tile surfaces twenty two of the twenty four JET divertor modules (i.e. 330° of the divertor) were vacuumed. The remaining two modules (one from octant 1 and the one from octant 5) were reserved for *post mortem* analysis and thus the tile surfaces were preserved. The surface area was scaled accordingly to take account of this. The sample from the divertor louvres was taken from 360°.

Vacuuming of the divertor tile surfaces (excluding those identified for *post mortem* analysis as discussed above) takes place during each JET intervention as part of the safety case for the shut-down procedures. The extent of the vacuuming is dependent on the level of refurbishment of the divertor. In brief the majority of the divertor tile surfaces were vacuumed in the 2004 and 2007 interventions, however the louvre area was only vacuumed in the 2004 intervention. To compensate for the differences in vacuuming history the dust samples were scaled with the total plasma seconds during different operating periods to give a scaled mass for the most recent 2007 - 2009 operation. So far the dust/flake samples collected from JET for quantitative analysis and characterisation have only come from the divertor where flaking deposits are found. Dust in tile gaps in the main chamber have not been sampled in JET.

Evaluation of the main chamber carbon source was also determined from spectroscopy using the CIII line at 465 nm with a horizontal line of sight onto the inner wall, in a region next to an inner wall guard limiter, sampling the inner wall. The total photon signal from the CIII line for the 2007 - 2009 operating period was evaluated. The spectroscopy results have also been scaled with the time for the non X-point (predominantly limiter) phases and for the X-point (divertor) phases to take account of the different erosion/deposition regimes during a plasma pulse. To provide the number of C atoms eroded from the main chamber the photon signal is scaled with the interaction area of the plasma, this has been estimated for the limiter phase as the wetted area of the plasma on the IWGLs and during the X-point phase the total plasma surface area. Finally an effective photon efficiency $S/XB = 3.08$ is applied. The effective photon efficiency is determined from calculation of main chamber fluxes given in [8].

3. **Results**

Erosion and deposition of IWGL tiles from the horizontal mid-plane and from the lower end of the inner limiter have been evaluated by profiling and cross section microscopy. The results indicate that a total of 1.75 g of carbon has been eroded from the mid-plane limiter tile pair (i.e. both left and right hand tiles) and 0.40 g has been deposited at the edge of the tile pair in the region beyond the last closed flux surface as shown in [9], resulting in net erosion of 1.35 g. For a tile pair lower down the inner limiter a net deposition of 0.30 g is observed. This is comprised of net deposition of 0.25 g with no erosion on the left tile (i.e. with plasma current) and net deposition of 0.05 g made up of 0.20 g of deposit and 0.15 g of erosion on the right tile. Similar results from IBA were observed in [9] whereby the left tile of the bottom limiter pair was dominated by deposition and the right tile was dominated by erosion. These IBA results also indicate that the reverse pattern of erosion and deposition is observed at the top of the limiter. Taking the result for both mid-plane and lower tile pairs, estimates for the total erosion, total deposition and net erosion at the IWGLs in the main chamber are estimated and summarised in Table 1. This assumes that the bottom three tile pairs and the top three tile pairs of the nineteen tile pairs making up an IWGL beam are subject to net deposition of 0.30 g and the remaining sixteen tile pairs are subject to net erosion of 1.35 g, as described above, and that there are sixteen IWGL beams in the JET vessel.

For an OPL tile pair (i.e. top and bottom tiles) also situated at the mid-plane of the main chamber there is a total deposition of 0.05 g and total erosion of 0.19 g giving a net erosion of 0.14 g. The net carbon source due to erosion from the outer limiter, assuming the result for this one tile and taking into account forty five OPL limiter pairs on each of twelve limiter beams is 72 g, Table 1. However the profiler results for this tile are subject to some errors due to the shape of the tile. The long thin tiles (345 mm x 26 mm) bound together in pairs are difficult to mount; there is tilt from the vertical and the sloping surface of the tiles increases errors in the profiling results. In order to assess this tile fully cross section microscopy of the deposits on the ends of the tiles is required. Based on secondary ion mass spectrometry of OPL tiles removed in 2007 it is known that a 1 µm W / 10 µm C coating was completely removed in the centre of the tiles and deposition of the order 2 µm was observed at the ends. This results in a total deposition of <0.01 g and total erosion of 0.24 g giving a net erosion of 0.23 g of carbon from the equivalent OPL pair. This equates to erosion for the whole vessel from the OPLs of 124 g, a factor of two higher than determined by profiling for the tiles removed in 2009.

Three dump plate tiles in the vessel from 2005 - 2009 were profiled. From the profiler results no strong erosion or deposition was observed from these tiles. On one of these tiles installed in 2005 there was a 1 μm W / 10 μm C marker coating stripe which has been almost completely eroded. It is still possible to observe where the marker stripe was located from the profilometry data but the metallic layer is not visible to the naked eye. This confirms that the level of erosion is of the order of 10 μm in the dump plate region. Assuming this level of erosion and using a scaling of 0.56 for the exposure period 2007 - 2009 based on total plasma time, and the dump plate area ~14 m$^2$, the upper limit for erosion from the dump plate region is 130 g.

Erosion for the IWC tiles, covering the inner vacuum vessel wall between the inner limiters, has also been calculated from the analysis of marker coatings by proton backscattering. The markers were exposed from 2005 - 2009 and the total C source is estimated at 230 g [7]. The fraction of this attributable to the 2007 - 2009 operating period is 129 g for the inner wall.

Based on the tile profiler measurements, cross section microscopy of cores taken from tiles and ion beam analysis, an upper estimate for the main chamber carbon source (erosion) for the 2007 - 2009 operating period is 436 g, Table 1.

Results for the total erosion, total deposition and net erosion/net deposition taking into account the number of tiles of each type found within the divertor are also shown in Table 1. Profiler results for Tiles 4, 6, 7 and the Load Bearing Tile (LBT) have been evaluated. Data for Tile 7 was for a tile installed in the vessel from 2005 - 2009 and has been scaled according to plasma time for the 2007 - 2009 operating period as for other tiles. Tiles 1, 3 and 8 have yet to be evaluated, however this will not significantly change the result as the deposition on the sloping surface of tiles 4 and 6 dominate the total amount of deposit of carbon in the divertor. For example Figure 3(a) shows deposit on Tile 6 is an order of magnitude thicker than the erosion observed on Tile 7, Figure 3(b). This is also supported by results presented in [12]. Based on the results available the net deposition onto tiles in the divertor is 533 g.

Although Tile 1 has not been analysed a thick deposit was observed on the top horizontal surface of Tile 1 at the inner divertor on a tile removed in 2007, this deposit had reached a critical thickness of ~120 μm and was spalling readily [10] therefore any carbon reaching this surface during the 2007 - 2009 operating period is more likely to contribute to the dust/flake sample taken from this region.

The masses of dust and flakes collected from the divertor region also contributes to the amount of carbon found in the divertor. The masses collected are shown in Figure 2. The largest amounts of dust collected were in the regions where significant deposition has taken place; notably at the inner divertor (115 g) where heavy deposition has been observed on the top horizontal surface of Tile 1s, from Tile 4s and Tile 6s (22 g and 51 g respectively) where thick deposits form on the sloping regions as shown in Figure 3(a) and also at the inner and outer louvres (91 g) remote from the plasma. In total the mass of dust/flakes collected after scaling for the areas surveyed and the vacuuming history of the tiles was 300 g.

The carbon source from the main chamber is compared with the CIII line spectroscopy from the mid-plane of the main chamber. The total carbon signal for the operating period (2007 - 2009) was $7.91 \times 10^{18}$ photons / $cm^2$ sr for 178449 s (49.6 hours). This is split into 44964 s (12.5 hours) of non X-point phase (predominantly limiter phase) and 133485 s (37.1 hours) of X-point phase. The signal is from a horizontal line of sight in the main chamber onto the inner wall near to an inner wall limiter and may therefore be lower than the signal expected directly from the inner limiter. The main issue in interpreting the main chamber source is establishing the appropriate area for scaling the raw data. During the limiter phase there is a strong interaction between the limiters and the plasma giving rise to a carbon source which is locally re-deposited in the main chamber. The integrated spectroscopy signal during the limiter phase is scaled with the *wetted* area of the IWGLs, this is determined as 3.6 $m^2$ from erosion zones visible on the IWGL tiles analysed and scaled to include all the IWGL tiles. This gives the mass of carbon eroded is calculated as 57 g during the limiter phase from CIII spectroscopy. This value is higher than the 35 g a total deposition (i.e., locally re-deposited carbon in the limiter phase) determined from the profiler measurements on the IWGL tiles. There are several sources of error in this calculation; (i) the evaluation of the wetted area for limiter-plasma interaction is difficult to define and therefore presents some error in the spectroscopy calculation, (ii) recycling of carbon can also lead to a higher carbon source being determined from spectroscopy signals and (iii) the determination of the limiter phase taken as all plasma time where an X-point has not been formed - this could include other events such as disruptions in the limiter phase calculation. Points (i) and (ii) could give rise to errors of up to 50%, whereas the inclusion of CIII signal from disruptions in the limiter phases (point (iii)) is expected to be negligible compared with the whole operating period under discussion.

The 28 g of locally re-deposited carbon observed on the OPL tiles from tile profiling data indicates that the interaction area of the plasma with the outer limiters during the limiter phases is less than that of the IWGLs. It is generally true that plasmas are established at the inner limiter rather than the outer limiters. However if the CIII signal were to take into account a wetted area from the OPLs then the carbon source from the main chamber would be even greater and would account for the locally re-deposited material on the outer limiter tiles.

It should be noted that although a strong interaction between the limiters and the plasma during the limiter phase is expected the total amount of erosion attributed to the limiter phase (i.e. assumed to equal the net deposition on the limiters) is only 20 - 30% of the total erosion found on the limiters for the entire operating period (i.e,. limiter and X-point phases). This either indicates (i) further erosion of the limiters occurs during the X-point phase or (ii) the amount of deposit found on the limiters as a result of local re-deposition *is not* directly representative of erosion during the limiter phase and that eroded material migrates further around the machine or to the divertor on X-point formation or (iii) the amount of locally re-deposited material *is* representative of erosion and has been under estimated by the analysis techniques. The first supposition is expected to be the case.

In order to determine the main chamber carbon source during the X-point phase from the spectroscopy signal the whole plasma area of 139 m$^2$ is used, giving ~2000 g of eroded carbon. During the X-point phase this mass of carbon will migrate from the main chamber into the divertor. This mass of carbon is a factor of 2-3 times higher than the net deposition observed in the divertor and the mass of dust/flakes collected, 533 g and 300 g respectively.

**4. Discussion**

The carbon balance of JET in the period 2007 - 2009 can be considered as the balance of *net erosion in the main chamber = net deposition in the divertor + dust/flake collected in the divertor + remote carbon*. The net erosion of the main chamber tiles has been evaluated from profiling and optical microscopy of tiles giving a value of 436 g. It is assumed that net eroded material from the main chamber will be deposited in the divertor during the X-point phase, whereas the difference between the gross and net eroded material will be locally re-deposited on the limiters during the limiter phase. Based on this assumption the net eroded material from the main chamber can be compared to the erosion sources determined using the spectroscopy signal during the X-point phase, which is ~2000 g. Deposition and dust/flakes in the divertor give a net deposition value of 833 g. Clearly there are some discrepancies arising in this balance. A comparison shows that the spectroscopic carbon source calculated for this operating period is somewhat higher than calculated for previous operating periods. For example the carbon source determined for the 2005 - 2007 operating period was 770 g [11]. A simple scaling of this value based on the total plasma time for the 2007 - 2009 operations which was 20% longer than the 2005 - 2007 operating period gives 924 g of carbon in the main chamber. This would be comparable with the net deposit and dust/flakes in the divertor but still a factor of two higher than the profiler results for the net erosion from the main chamber. If the carbon source estimated from spectroscopy in the main chamber is accurate then the *remote* carbon is in the range 100 g -1100 g. Such remote carbon could be; (i) directly pumped from the machine as $CD_4$, (ii) be in remote areas as flakes under the divertor as seen during earlier JET shutdowns or, (iii) possibly held as long chain hydrocarbon on cryo-panels. The evaluation of these potential sources of remote carbon is not undertaken here as there are no quantitative data. However the contribution of remote carbon is unlikely to exceed the other errors contributing to the balance. In fact carbon pumped directly from the TEXTOR vessel is estimated as low as ~5% of the erosion source [2].

As far as the net deposit in the divertor is concerned the value obtained from profiling and the amount of dust collected (833 g) are lower than for earlier campaigns showing a total of 1700 g for 1999 - 2001, [4]. It is also slightly lower than shown in [12] where significant levels of deuterium observed in these same tiles 4 and 6 scale to give a mass of 1370 g of carbon deposit for an average D/C ratio of 0.5. The deposition rates in the divertor are also lower, $6x10^{-3}$ g/s compared with $35x10^{-3}$ g/s for 1999- 2001 [4].

The errors associated with this calculation of net erosion and deposition arise from the scaling up from representative tiles to the whole chamber, the density of the deposits and the CFC bulk used in calculation of the mass of material and the sensitivity of the tile profiler.

The tiles analysed from the main chamber represent < 1% of the total surface area and the tiles analysed in the divertor represent ~1% of the surface area in the divertor. Toroidal symmetry is accepted therefore the results from individual divertor tiles are scaled to the whole divertor. It is poloidal variations in the main chamber that could make a significant difference when scaling, particularly in the case of the limiter tiles. If all nineteen tile pairs making up an inner limiter were assumed to show the same level of erosion as the mid-plane tile pair then the total erosion would increase by 20%. Conversely if the belt of erosion around the mid-plane of the vessel was assumed to be narrower by an additional three rows of limiter tiles then the erosion would decrease by 10%.

An additional source of error is the choice of density for the deposited carbon layers. The density of deposited carbon may be lower (0.8 g/cm$^3$ [6]) than used in these calculations, 1 g/cm$^3$ [5]. This would increase the net erosion from the main chamber to 485 g (an increase of 10%) and decrease the net deposition in the divertor to 412 g (a decrease of >20%), bringing the balance closer. In fact it is likely that the density of re-deposited carbon varies depending on where it is found in the machine and the operating conditions. Clearly the variation in density of the deposited material has an effect on the overall picture of the carbon balance.

These results indicate that the net erosion source from the main chamber determined from profiling may be underestimated. In order to account for the discrepancy (200 - 300 g) an additional erosion depth of >30 μm would need to be eroded across the main chamber (assuming 10 m$^2$ interaction area). This amount of erosion would be resolvable from the profiler measurements as variations of ± 10 μm (equivalent to 100 g of carbon) can be detected. Therefore errors from tile profiling data could account for 33 - 50% of the discrepancy.

Whilst individually these errors do not compensate for the discrepancy between the main chamber erosion and divertor deposition a combinations could account for a significant amount of the 200 - 300 g difference.

Another factor of interest from this analysis is the conversion factor from deposit to dust/flakes. Based on the results for total deposition in the divertor the conversion rates are up to 36%. The conversion factor is likely to vary depending on the thickness and stability of the deposits formed. For example from the inner vertical divertor tiles, including the heavily deposited horizontal surface on Tile 1, 115 g of dust/flakes were collected. The deposit on this surface reaches its critical thickness of 120 μm in a typical operating period and does not increase when tiles are left in for more than one operating period [10]. Therefore for the majority of tiles in the tile 1 location a conversion factor of 100% with a calculated mass of 54 g of carbon expected to spall from this horizontal surface during the 2007 - 2009 operating period. Evidence that the growth and spallation of carbon have reached an equilibrium is also found on the inner and outer divertor corner tiles 4 and 6 that have been in JET from 2005 - 2009, i.e., two operational periods. When the total amount of deposit on these two tiles determined from profiling is scaled by total plasma seconds for the 2007 - 2009 operating period the amount of deposit attributed to this period is 20% lower than that determined for tiles that were new in the vessel in 2007. This indicates that the deposits have reached there critical thickness and layers are readily spalling.

The dust/flakes collected from Tiles 4 and 6 in the vessel from 2007 - 2009 indicate a conversion factor of 7% and 19% respectively. The overall conversion rate of 36% is therefore reasonable, as the areas covered by Tile 4 and 6 is larger than for the top of Tile 1. In contrast the mass of dust/flakes collected from the vertical outer tiles (Tile 7 and 8) is < 1g, indicative of this region of the divertor being an erosion zone. However the mass of dust/flakes collected may also be revised down following analysis by as much as 20%, particularly if foreign objects introduced from shutdown activities are found in the samples. This would decrease the conversion factor.

5. **Conclusion**

The main chamber net erosion (436 ± 100 g) and divertor net deposition plus dust/flakes (up to 833 g) determined experimentally agree within a factor of two. After considering the errors in scaling discussed in section 4 an agreement within a factor of two is acceptable. Further evaluation of the profiling data by comparison with cross sectional microscopy data is on going and may bring the balance closer together. The main chamber carbon source from spectroscopy is estimated to be as high as 2000 g. This is somewhat larger than for previous campaigns [4]. An explanation for these differences has not been identified. In addition it has not been possible to provide quantitative data on remote carbon which may contribute to complete the global picture. Whilst the carbon balance is not exact the results continue to support the picture of carbon eroded in the main chamber and being transported via the scrape off layer into the divertor. These results provide a benchmark for comparison of erosion and deposition between the *all carbon* and *all metal* scenarios in JET. Beryllium and tungsten main chamber tiles and tungsten divertor tiles from the "all metal" ILW wall will be removed during the 2012 JET intervention and analysis will start in 2013.

6. Acknowledgements

This work, part-funded by the European Communities under the contract of Association between EURATOM/CCFE was carried out within the framework of EFDA. The views and opinions expressed herein do not necessarily reflect those of the European Commission. This work was also part-funded by the RCUK Energy Programme under grant EP/I501045.

8. **Tables**

|  | Total Deposition (g) | Total Erosion (g) | Net erosion (-) / Net deposition (+) (g) |
|---|---|---|---|
| Inner Wall Guard Limiter | 35 | -175 | -140 |
| Outer Poloidal Limiter | 14 | -50 | -36 |
| Dump plate | - | -130 | -130 |
| Inner Wall Cladding* | - | - | -129 |
| Tile 4 | 312 | 0 | +312 |
| Load Bearing Tile | 21 | -38 | -17 |
| Tile 6 | 272 | <-1 | 272 |
| Tile 7 | 1 | -36 | -35 |

Table 1 Summary of total erosion, total deposition and net erosion/deposition estimated for the whole vessel from profiling of main chamber and divertor tiles. * Data from ion beam analysis [7].

## 9. Figure Captions

Figure 1 Poloidal cross section of the JET vessel with the MkII-HD divertor. The results presented in this paper are indicated and come from Tile 1, 4, Load Bearing Tile, Tile 6 and 7 in the divertor and tiles from the Inner Wall Guard Limiter, Dump Plate region and the Outer Poloidal Limiter in the main chamber. Inner Wall Cladding tiles are recessed behind the IWGL and fill the space between the sixteen IWGL beams.

Figure 2 Cross section of Mk-HD divertor section showing the mass of dust/flakes vacuumed from different regions.

Figure 3 Profiler results for (a) Tile 6 (b) Tile 7.

## 10. Figures

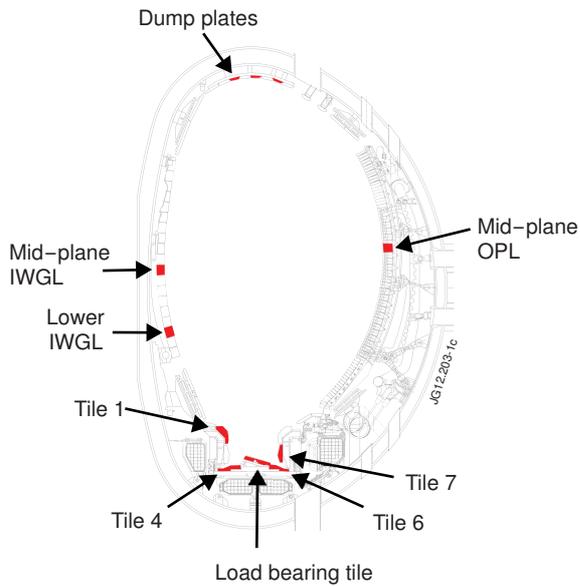

Figure 1 Poloidal cross section of the JET vessel with the MkII-HD divertor. The results presented in this paper are indicated and come from Tile 1, 4, Load Bearing Tile, Tile 6 and 7 in the divertor and tiles from the Inner Wall Guard Limiter, Dump Plate region and the Outer Poloidal Limiter in the main chamber. Inner Wall Cladding tiles are recessed behind the IWGL and fill the space between the sixteen IWGL beams.

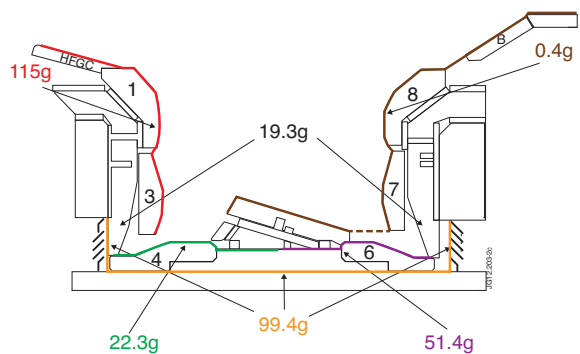

Figure 2 Cross section of Mk-HD divertor section showing the mass of dust/flakes vacuumed from different regions.

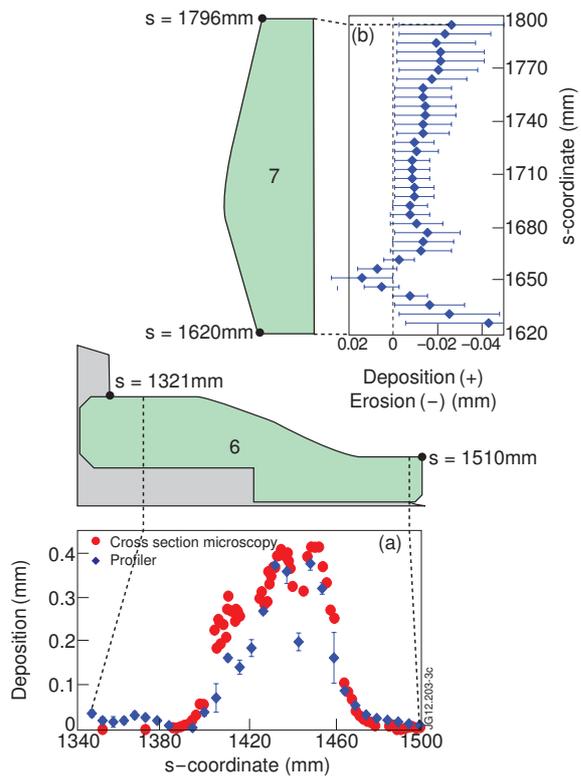

Figure 3 Profiler results for (a) Tile 6 (b) Tile 7.